\documentstyle[prd,aps]{revtex}

\begin{document}

\draft

%
%

\title{No-scalar hair conjecture in asymptotic de-Sitter 
spacetime\thanks{TIT/HEP-403}}

\author{Takashi Torii\thanks{electronic
mail:torii@th.phys.titech.ac.jp} and
Kengo Maeda\thanks{electronic 
mail:maeda@th.phys.titech.ac.jp}}
\address{Department of Physics, Tokyo Institute of 
Technology,
Meguro-ku, Tokyo 152-0033, Japan}

\author{Makoto Narita\thanks{electronic
mail:narita@se.rikkyo.ac.jp}}
\address{Department of Physics, Rikkyo University,
Toshima-ku, Tokyo 171-8501, Japan}

\date{\today}

\maketitle

\begin{abstract}
We discuss the no-hair conjecture in the
presence of a cosmological constant. For the first
step  the real scalar field is considered as the matter
field and the spacetime is assumed to be static spherically
symmetric. If the scalar field is massless or has a convex 
potential such as a mass term, it is proved that there is no
regular black hole solution. 
For a general positive potential,
we search for black hole solutions which support the 
scalar
field with a double well potential, and find them by
numerical calculations. 
The existence of such solutions depends on the values
of the vacuum expectation value and the self-coupling 
constant
of the scalar field. When we take the zero horizon radius limit,
the solution becomes a boson star like solution which we 
found
before. However new solutions are found to be unstable 
against the linear perturbation. As a result we can conclude 
that the no-scalar hair
conjecture holds in the case of  scalar fields with a convex or 
double well potential.
\end{abstract}


%
%
%

\section{Introduction}
\label{sec:Introduction}

Recent developments of observational techniques  have increased 
interest about  black holes and increasing
data shows  evidence of  supermassive black holes
in the center of galaxies, and solar mass size black holes
which form binary systems.  
From the theoretical view point, however,
there remain many unsolved fundamental issues related to the
black holes.
One of them is the validity of the black hole no-hair
conjecture proposed by Ruffini and Wheeler\cite{RW}. The 
black hole
no-hair conjecture states that after the gravitational 
collapse
of the matter field, the resultant black hole approaches 
stationary
spacetime and that all its multipole moments are then 
uniquely
determined by two parameters, $M$ and $a$, which are 
physically interpreted as the mass 
and the angular momentum of the black 
hole.
When the source has a net charge $Q$, then of course the 
parameter 
is also requested to uniquely determine its (electric and
gravitational) multipole moments. The black hole no-hair
conjecture is supported by the black hole uniqueness 
theorems
in electrovacuum theories\cite{BHT}  and by the works of
Chase\cite{Cha}, Bekenstein\cite{Bek},
Hartle\cite{Har} and Teitelboim\cite{Tei} which show that
stationary black hole solutions are hairless in a variety of
theories coupling classical fields to Einstein gravity.

However a non-trivial static, spherically symmetric solution, 
called
colored black hole, was discovered in the Einstein-Yang-
Mills 
system\cite{CBH}. It can be interpreted as the self-gravity
of the Yang-Mills field being supported by its repulsive force. 
Although
this solution is found to be unstable both
in the gravitational sector\cite{GRA} and in the sphaleron
sector\cite{SPH},  non-Abelian hair is generic and many 
other 
non-Abelian black holes were discovered after the colored
black hole. Note that some of them are stable  and that such 
solutions
are the counterexamples of the black hole no-hair 
conjecture.
The essential difference between the field equations of
Einstein-Yang-Mills systems  and those of the
Einstein-Klein-Goldon system is in the form of the potential 
term.
By using these properties, Bekenstein\cite{Bek,Bek2} and
Sudarsky\cite{Sud} provided simple proofs for the no-scalar
hair theorem in spherically symmetric 
asymptotically flat spacetime, in the case
where the matter consists of a single scalar field with a 
convex
potential and in the extended case where the matter consists 
of 
multiple scalar fields with an arbitrary positive potential.
Heusler also proved the no-scalar hair theorem by using 
a scaling technique\cite{Heu}.

Most of the proofs of the no-hair theorems impose, however,  
asymptotic flatness.
Hence the following natural question arises:
{\it Can we extend no-hair theorems to spacetimes with 
different 
asymptotic structure?} This is the main issue of this paper.
Here we consider the system including the cosmological 
constant
and only one real scalar field as a matter field for the 
first step.
The importance of the cosmological constant
comes not only  from the theoretical aspect but also from 
the observational
results of our universe. For example some astrophysicists 
have
pointed out that the small cosmological constant may explain
the observed number count of galaxies\cite{NCG}, and the 
recent observation
of the type Ia supernova at large redshift  also supports 
the 
universe with a cosmological constant\cite{OBS2}. 
Furthermore, in the 
early universe, we usually expect a vacuum energy, which is 
equivalent to the cosmological constant.
As for the black hole solutions with a cosmological
constant, the family of the Kerr-Newman-de Sitter solutions
are known as the exact solutions. Besides these the cosmic 
colored
black hole solution is derived\cite{CCH}. These solutions 
are
an extension of the Kerr-Newman solutions and the colored black 
hole
to the non-zero cosmological constant case. Interestingly,
although there are black hole solutions in the Einstein-
Maxwell-dilaton
system\cite{EMD}, no black hole solution exists if we take 
the
cosmological constant into account\cite{EMDL}.
Hence the cosmological constant can strongly affect the 
existence of
black hole solutions.

If we find the physical solutions with scalar hair, the 
no-scalar 
hair conjecture
can not hold in asymptotic de Sitter spacetime 
and such solutions may 
have an 
influence on the cosmology. If there are
no such solutions, we can say that the no-scalar hair 
conjecture 
holds in our system. In other words, 
the weak cosmic no-hair conjecture holds in the sense that 
all
initially expanding universes with positive cosmological
constant approach the ``de Sitter spacetime"
asymptotically, except for locally distributed black holes. 
\cite{CNH}.
Hence it must be stressed that the black hole no-hair 
conjecture
in asymptotic 
de Sitter spacetime is strongly related to the weak
cosmic no-hair conjecture.

This paper is organized as follows. 
In section II we introduce the model and the basic equations 
and 
try to extend the no-hair theorem in a spherically symmetric 
system.
In section III we derive black hole solutions numerically 
in the model that
the real scalar field has a double well potential and
discuss their properties. We investigate the stability
of these solutions in section IV. We give our conclusions 
and 
remarks in the final section.

%
%
%

\section{Model and Basic Equations}
\label{sec:Model}

We will consider the model given by the action
\begin{equation}
S=\int d^4x \sqrt{-g} \left[ \frac1{16\pi G} (R-2\Lambda)
   -\frac12 (\nabla \phi)^2 
    -V(\phi) \right],
\label{action}   
\end{equation}
where $\phi$ is the real scalar field and $V(\phi)$ is its 
potential. We shall assume a spherically symmetric spacetime 
and adopt the Schwarzschild type metric,
\begin{equation}
ds^2 = -\left(1-\frac{2Gm}{r} -\frac{\Lambda}{3} r^2\right) 
  e^{-2\delta} dt^2
  +\left(1-\frac{2Gm}{r} -\frac{\Lambda}{3} r^2\right)^{-1} 
dr^2
  +r^2 \left(d\theta^2 + \sin^2 \theta d\varphi^2 \right).
\label{metric}   
\end{equation}
The mass function $m$
and the lapse function
$\delta$ depend on both the time coordinate $t$ and the
radial coordinate $r$. The mass function is the quasilocal 
mass defined by Ref. \cite{QM}, which is the 
gravitational energy subtracted by the energy due to the
cosmological constant $\Lambda$. In other words, it is 
considered
as the energy of the matter field.
In spherically symmetric spacetime, the energy $m$
is nondecreasing in the outgoing null or spacelike 
direction in the region $1-2Gm/r -3\Lambda r^2 >0$
if the matter fields satisfy the dominant energy
condition.

Varying the action (\ref{action}) and substituting 
{\it Ansatz}
(\ref{metric}), we derive the field equations
\begin{equation}
\tilde{m}^{\prime} = 4\pi \tilde{r}^2 
    \left\{ \frac{1}{2} \left(1-\frac{2\tilde{m}}{\tilde{r}}
    -\frac{1}{3}\tilde{r}^2\right)^{-1} e^{2\delta}
    \dot{\tilde{\phi}}^{2}
    +\frac{1}{2} \left(1-\frac{2\tilde{m}}{\tilde{r}}
    -\frac{1}{3}\tilde{r}^2 \right)
    \tilde{\phi}^{\prime 2}
    + \tilde{V}(\tilde{\phi}) \right\},
\label{beq1}   
\end{equation}
\begin{equation}
\delta^{\prime} = -4\pi \tilde{r} 
       \left[ \left(1-\frac{2\tilde{m}}{\tilde{r}}
    -\frac{1}{3}\tilde{r}^2 \right)^{-2} e^{2\delta}
    \dot{\tilde{\phi}}^{2}+\tilde{\phi}^{\prime 2} \right],
\label{beq2}   
\end{equation}
\begin{equation}
\dot{\tilde{m}} = 4\pi \tilde{r}^2 
     \left(1-\frac{2\tilde{m}}{\tilde{r}}
    -\frac{1}{3}\tilde{r}^2\right) 
    \dot{\tilde{\phi}} \tilde{\phi}^{\prime},
\label{beq3}   
\end{equation}
\begin{equation}
-\left[e^{\delta} \left(1-\frac{2\tilde{m}}{\tilde{r}}
   -\frac{1}{3}\tilde{r}^2 \right)^{-1} \dot{\tilde{\phi}} 
\right]^{\cdot}
+\frac{1}{\tilde{r}^2} \left[\tilde{r}^2 e^{-\delta}
   \left(1-\frac{2\tilde{m}}{\tilde{r}}
   -\frac{1}{3}\tilde{r}^2\right) \tilde{\phi}^{\prime} 
\right]^{\prime}
=e^{-\delta} \frac{d\tilde{V}(\tilde{\phi})}{d\tilde{\phi}}.
\label{beq5}   
\end{equation}
Here, we have used the dimensionless variables, 
$\tilde{t} \equiv \sqrt{\Lambda}t$,
$\tilde{r} \equiv \sqrt{\Lambda}r$, 
$\tilde{m} \equiv \sqrt{\Lambda}Gm$,
$\tilde{\phi} \equiv \sqrt{G}\phi$ 
and 
$\tilde{V} \equiv GV/\Lambda$.
A dot and a prime in the field equations 
denote derivatives
with respect to $\tilde{t}$ and $\tilde{r}$, respectively.
In this section and the next section
we consider only the   static solutions, hence 
we drop the time derivative term of the
field equations for a while.


For the boundary conditions of the metric functions on the
black hole event horizon (BEH) and on the 
cosmological event horizon (CEH), we impose the following
three {\it Ans\"atze}.

(i) The existence of a regular BEH $r_{B}$; i.e.,
\begin{eqnarray}
2Gm(r_{B}) & = & r_{B}
        \left(1-\frac{\Lambda}{3}r_{B}^2 \right),  
\label{bcondi3}   \\
\delta(r_{B}) & < & \infty.  \label{bcondi4}   
\end{eqnarray}

(ii) The existence of a regular CEH $r_{C}$; i.e.,
\begin{eqnarray}
2Gm(r_{C}) & \equiv & 2GM  =
r_{C} \left(1-\frac{\Lambda}{3}r_{C}^2 \right),  
\label{bcondi1}   \\
\delta(r_{C}) & < & \infty.  \label{bcondi2}   
\end{eqnarray}
Here we assume $\delta(r_{C})=0$. 
Note that if we are interested in a different boundary 
condition such as
$\delta \to \delta^{\ast} \ne 0$, we can always have such
a boundary
condition without further calculation  by rescaling 
the time coordinate. 
That is,
introducing
$\bar{\delta} \equiv \delta - \delta^{\ast}$, and rescaling 
the
time coordinate as 
$\bar{t}=e^{-\delta^{\ast}} \tilde{t}$,
we recover our boundary condition.

(iii) The non-existence of singularity between 
BEH and CEH; i.e., for $r_{B} < r <r_{C}$,
\begin{equation}
2Gm(r)  < r \left(1-\frac{\Lambda}{3}r^2 \right).  
\label{bcondi5}  
\end{equation}
As for the scalar field, we impose the finiteness of
itself and its derivatives. These conditions guarantee
that the curvature invariant 
$I=R_{\mu\nu\rho\sigma}R^{\mu\nu\rho\sigma}$
is finite and that no naked singularity appears.

Now we investigate whether or not the no-scalar
hair theorems obtained by Bekenstein\cite{Bek,Bek2} and 
Sudarsky\cite{Sud}
can be extended to our system.
First we consider the case of the
scalar field with convex potential,
e.g. $V(\phi) = m_{\phi}^2 \phi^2$. 
In the static spacetime the equation of the scalar field is
\begin{equation}
\left(\tilde{r}^2 e^{-\delta}f 
  \tilde{\phi}^{\prime} \right)^{\prime}
=\tilde{r}^2e^{-\delta} 
\frac{d\tilde{V}(\tilde{\phi})}{d\tilde{\phi}},
\label{beq6}   
\end{equation}
where $f = 1-2\tilde{m}/\tilde{r}-\tilde{r}^2/3$.
Multiplying by $\tilde{\phi}$ and integrating from BEH to 
CEH,
we obtain
\begin{eqnarray}
\int^{\tilde{r}_C}_{\tilde{r}_B} d \tilde{r} & &
 \left\{ \tilde{r}^2e^{-\delta} \tilde{\phi} 
  \frac{d\tilde{V}(\tilde{\phi})}{d\tilde{\phi}}
 - \tilde{\phi} \left(\tilde{r}^2 e^{-\delta} f 
  \tilde{\phi}^{\prime} \right)^{\prime} \right\} 
\nonumber \\
& & =\int^{\tilde{r}_C}_{\tilde{r}_B} d \tilde{r} 
 \tilde{r}^2e^{-\delta} \left(\tilde{\phi} 
  \frac{d\tilde{V}(\tilde{\phi})}{d\tilde{\phi}}
 + f \tilde{\phi}^{\prime 2} \right)
  - \left[\tilde{r}^2 e^{-\delta} f
  \tilde{\phi} \tilde{\phi}^{\prime} 
  \right]^{\tilde{r}_C}_{\tilde{r}_B}
  \nonumber \\
& & =\int^{\tilde{r}_C}_{\tilde{r}_B} d \tilde{r} 
 \tilde{r}^2e^{-\delta} \left(\tilde{\phi} 
  \frac{d\tilde{V}(\tilde{\phi})}{d\tilde{\phi}}
 + f \tilde{\phi}^{\prime 2} \right) =0
  \label{noh1}
 \end{eqnarray}
Since $\tilde{V}$ is a convex potential, the integrand is 
positive
semi-definite. Hence the possible solution is $\tilde{\phi} 
\equiv 0$.
As a result the same argument as Bekenstein's holds and the
no-scalar hair theorem is complete in the convex potential
case\cite{CJ}. When the scalar field is massless, i.e.,
$\tilde{V}\equiv 0$, Eq. ({\ref{noh1}) also
demands $\tilde{\phi} \equiv 0$, which is the no-hair theorem 
for a
massless scalar field.

Next we proceed to the theorem proved by Sudarsky.
First we rewrite Eq.~(\ref{beq5}) in the static spacetime as
\begin{equation}
\left(e^{-\delta} f \tilde{\phi}^{\prime} \right)^{\prime}
+\frac2{\tilde{r}} e^{-\delta} f \tilde{\phi}^{\prime}
-e^{-\delta} 
\frac{d\tilde{V}(\tilde{\phi})}{d\tilde{\phi}}=0.
\end{equation}
Multiplying $\tilde{\phi}^{\prime}$ and integrating once, 
we 
obtain
\begin{equation}
E^{\prime} = -a e^{-\delta} \tilde{\phi}^{\prime 2},
\label{EE}
\end{equation}
where 
\begin{eqnarray}
E & = & e^{-\delta} \left( f \tilde{\phi}^{\prime 2}
   -\tilde{V} \right),  \\
a & = & \frac{2}{\tilde{r}} 
  \left(1-\frac{3\tilde{m}}{2\tilde{r}}
   -\frac{1}{2}\tilde{r}^2 \right).
\end{eqnarray}
In the $\Lambda=0$ case, i.e., asymptotically flat case,
$E=-\tilde{V}<0$ on BEH, and
$a$ is always positive between BEH and CEH
because of the regularity.
(Note that we have normalized variables by
$\Lambda$, $a = 2(1-3m/2r)/r$
in the $\Lambda=0$ case.)
Hence $E$ decreases monotonically and then becomes negative
everywhere.
However $E$ must approach zero in the asymptotic 
region\cite{Sud}. 
This means that there is no regular solution in this system.
On the other hand, in the $\Lambda \ne 0$ case, 
$E=-\tilde{V}<0$ on both horizons and $a$ is no longer
positive definite. Hence the inconsistent behavior of $E$ 
can not
be derived from Eq. (\ref{EE}).
Furthermore asymptotic behavior does not reject the scalar
hair in the arbitrarily positive potential case\cite{CJ}.

This suggests that the method adopted here may not 
be appropriate 
to prove the black hole no-hair or
that the black hole solutions
with non-trivial scalar hairs may exist.
We will search for such  solutions in the next section by using
numerical calculations.

%
%
%

\section{Black Hole Solutions with Scalar Hair}
\label{sec:Solutions}

In this section we investigate the system
including a double well potential $V(\phi) = 
\lambda(\phi^2 -v^2)^2/4$ as an example of a positive
potential. In this case there are two
trivial black hole solutions. One is 
(a) $\tilde{\phi} \equiv \tilde{v}$, $\tilde{m} 
\equiv \tilde{M}$, 
$\delta \equiv 0$.
The scalar field  takes its vacuum expectation value 
and does not contribute to the spacetime at all.
This is the usual 
Schwarzschild-de Sitter solution.
The other is 
(b) $\tilde{\phi} \equiv 0$, 
$\tilde{m} = \tilde{M} + \pi \tilde{\lambda} 
\tilde{v}^2 \tilde{r}^3/3$, 
$\delta \equiv 0$, 
where $\tilde{\lambda} \equiv \lambda/G \Lambda$ and
$\tilde{v} \equiv \sqrt{G}v$.
The scalar field sits on the top of the potential
barrier everywhere and its contribution to the spacetime
can be interpreted as the 
``effective cosmological constant" 
defined by $\Lambda_{eff} = 2\pi\lambda \tilde{v}^4$.
We call this solution  the excited 
Schwarzschild-de Sitter solution.

The equation of the scalar field (\ref{beq5}) is rewritten 
as 
\begin{equation}
f \tilde{\phi}^{\prime \prime}
+ \left[\left(\frac2{\tilde{r}} +4\pi \tilde{r} 
\tilde{\phi}^{\prime 2}
   \right)f
   + f^{\prime}\right] \tilde{\phi}^{\prime}
= \tilde{\lambda}\tilde{\phi}\left(\tilde{\phi}^2 
-\tilde{v}^2\right).
\label{beq4-2}   
\end{equation}
On the event horizon Eq. (\ref{beq4-2}) becomes
\begin{equation}
   f^{\prime}\tilde{\phi}^{\prime}
= \tilde{\lambda}\tilde{\phi}\left(\tilde{\phi}^2 
-\tilde{v}^2\right).
\label{beq4-3}   
\end{equation}
Since $f'$ is positive on the BEH, $\phi'(r_{B})>0$
if $-v < \phi(r_{B}) <0$, $v<\phi(r_{B})$, and
$\phi'(r_{B})<0$
if $\phi(r_{B})<-v$, $0<\phi(r_{B})<v$. 
On the other hand
$f'$ is negative on the CEH, $\phi'(r_{C})>0$
if $\phi(r_{C})<-v$, $0<\phi(r_{C})<v$
and $\phi'(r_{C})<0$
if $-v < \phi(r_{C}) <0$, $v<\phi(r_{C})$.
At the extremum point of $\phi$, Eq. (\ref{beq4-2}) becomes
\begin{equation}
f \tilde{\phi}^{\prime \prime}
= \tilde{\lambda}\tilde{\phi}\left(\tilde{\phi}^2 
-\tilde{v}^2\right).
\label{beq4-4}   
\end{equation}
Since $f$ is always positive between BEH and CEH, $\phi$ has
no local maximum if $-v < \phi(r) <0$, $v<\phi(r)$
and $\phi$ has no local minimum if 
$\phi(r)<-v$, $0<\phi(r)<v$. Now we estimate
the possible behavior of the scalar field from the
above constraints. We can restrict $\phi(r_{B})>0$
without loss of generality because the
scalar field has reflection symmetry. If
$\phi(r_{B})>v$, $\phi$ increases around the BEH. Since
there is no maximum point $\phi>v$, it continues to 
increases monotonically to the CEH. 
However on the CEH $\phi^{\prime}$
must be negative. This is a contradiction.
When $0<\phi(r_{B})<v$, $\phi$ decreases around the
BEH. Since there is no minimum in this region, $\phi$
must pass over the potential barrier ($\phi=0$).
Continuing to decrease beyond $\phi=-v$,
$\phi$ can not satisfy the boundary condition on the CEH
because $\phi$ has no minimum in $\phi<-v$.
Hence $\phi$ must stop at a certain value in $-v<\phi<0$
or have a minimum. In the latter case $\phi$ passes over 
the potential barrier again.
In this way $\phi$ may oscillate any number of times and
takes a value $-v <\phi <v$ on the CEH. For each
oscillation $\phi$ has to go over the potential barrier.

Now we search for non-trivial static solutions using 
numerical
analysis. We drop the time derivative term of the field 
equations  
(\ref{beq1}), (\ref{beq2}) and (\ref{beq5}) and
integrate them from the BEH with the boundary
conditions (\ref{bcondi3}) and (\ref{bcondi4}). Since
the equation of the scalar field  (\ref{beq5}) becomes 
singular on 
the event horizons, we expand the equation and variables by
power series of $\tilde{r}-\tilde{r}_{B}$ to guarantee
the regularity on the BEH, and use their analytic solutions
for the first step of integration. For most of the 
values
of $\tilde{\phi} (\tilde{r}_{B})$, the scalar field diverges
as the integration approaches the CEH. Hence we have to 
find a 
suitable value of $\tilde{\phi} (\tilde{r}_{B})$ in order to
satisfy the boundary conditions (\ref{bcondi1}) and 
(\ref{bcondi2}).
In this sense $\tilde{\phi} (\tilde{r}_{B})$ is a shooting 
parameter.

We found the non-trivial solutions when $\tilde{\lambda}$, 
$\tilde{v}$ and $\tilde{r}_{B}$ satisfy a certain condition 
which will discussed later.
They are classified into several families by the node
number $n$ of the scalar field. 
Configurations of the field functions of the
solutions with $\tilde{\lambda}=700$, $\tilde{v}=0.1$ and 
$n=1$ for different radii of BEH are shown in Fig.~1.
The left and right end points of each solid line 
are BEH and CEH, respectively. A dotted line traces them.
A dashed line is the regular solution, which 
can be considered as the solutions
in $\tilde{r}_{B}\to 0$ limit~\cite{TT1}.
The structures of the new solutions do not concentrate 
around the BEH but
spread out to a cosmological scale.
As the BEH becomes large, the CEH becomes small and the 
values of the scalar field on each event horizon approaches 
its vacuum value $\tilde{\phi}= \pm \tilde{v}$. Hence the 
scalar field must vary rapidly in a small region and 
become
steep. However, as the BEH becomes even larger, 
$\tilde{\phi}(\tilde{r}_{B})$ and 
$\tilde{\phi}(\tilde{r}_{C})$ 
approach zero 
and finally
BEH and CEH coincide to be the extremal solution. 

We show the $\tilde{\lambda}$ dependence of the scalar
field for $\tilde{r}_B=0.2$, $\tilde{v}=0.1$ and 
$n=1$ in Fig. 2. For large $\tilde{\lambda}$, which
can be considered as the case that the cosmological constant
is small, the scalar field remains at its 
vacuum value $\tilde{\phi} \approx \tilde{v}$ even
for quite large $\tilde{r}$ around the BEH. As $\tilde{\lambda}$
decreases, the amplitude of $\tilde{\phi}$
becomes small and finally the solution coincides
with the excited Schwarzschild-de Sitter solution (b)
at a certain critical value $\tilde{\lambda}_{min}\cong 
354.8$.
These properties are qualitatively the same as the regular
solution discussed in Ref.~\cite{TT1}.
For  different values of parameters 
$\tilde{v}$, we obtain  similar results to those 
mentioned above.

We show the $\tilde{M}-\tilde{r}_{B}$ diagram of new 
solutions in Fig. 3, where $\tilde{M}$ is the quasi-local 
mass
on the CEH.
We fix $\tilde{v}=0.1$ and $n=1$.
We also show (a) the Schwarzschild-de Sitter branch
and (b) the exited Schwarzschild-de Sitter branch by
dashed  and dotted lines, respectively, for comparison.
We find that each solution branch turns at $\tilde{M} =1/3$.
From the boundary condition (\ref{bcondi1})
\begin{equation}
M =  \frac{r_{C}}{2G} \left(1 -\frac{\Lambda}3 {r_{C}}^2 
\right).
\label{mass}   
\end{equation}
Since the BEH has a one to one correspondence to the CEH around the
turning point,
\begin{equation}
\frac{\partial M}{\partial r_{B}}
= \frac{\partial r_{C}}{\partial r_{B}} 
\frac{\partial M}{\partial r_{C}}
= \frac12 \frac{\partial r_{C}}{\partial r_{B}}
    \left(1- \Lambda r_{C}^2 \right) =0,
\end{equation}
on the turning point.
From Fig.~3, $\partial r_{C}/\partial r_{B}$ does not 
vanish, 
hence $\tilde{r}_{C} = \sqrt{\Lambda} r_C =1$
at the turning point. This means $\tilde{M} =1/3$.

The properties of the new solution branch are quite 
different depending on $\tilde{\lambda}$.
For small
$\tilde{\lambda}$ there is only one solution for each
BEH radius. As the BEH radius increases, the branch 
approaches  the 
excited Schwarzschild-de Sitter branch and finally both
branches coincide at the maximum BEH 
radius.
At this point the
BEH and CEH coincide and the solution becomes extremal.
There is the minimum BEH radius, below which
only trivial solutions, i.e.,  Schwarzschild-de Sitter 
solutions 
(a) and (b)
exist. 
The non-trivial branch is very close
to the excited Schwarzschild-de Sitter branch. As we can see 
in Fig.~2,
the scalar field remains near the top of the potential
barrier ($\phi \sim 0$) for small $\tilde{\lambda}$. 
The main contribution to $M$
must be this potential energy of the scalar field. 
Hence the solution is similar to the 
excited Schwarzschild-de Sitter solution.

On the other hand, for large $\tilde{\lambda}$, there are two
solutions, which have the same BEH radius.
This only occurs for a narrow range of radii
The maximum BEH radius is not at the extremal point
but a turning point plotted by dots in Fig. 3.
On the small BEH radius , 
we can take the
$r_{B} \to 0$ limit even for the non-trivial branch, 
and the solution becomes the regular 
solution\cite{TT1}.
From Fig.~2 the scalar field takes an almost vacuum value near 
both horizons and passes rapidly on the top of the potential
barrier.
As a result the non-trivial branch is quite different from
the excited Schwarzschild-de Sitter solution.

In Fig.~4 we show the critical parameters in the
$\tilde{\lambda}-\tilde{v}$ 
plane for $r_{B}=0.6$ and $n=1$. There is a solution only in 
the
region between $\tilde{\lambda}_{mim}$ and 
$\tilde{\lambda}_{max}$. The existence of the 
critical line $\tilde{\lambda}_{min}$ 
plotted by the dashed line
is due to 
reasons similar to those in the regular solution case. 
In the regular solution case the critical lines are
derived by investigating the perturbation from the
de Sitter solution and are expressed
as
$( \tilde{R}_{cos}/\tilde{\lambda}_{Comp} )^2
= n(2n+3)$,
where $\tilde{R}_{cos}=
\sqrt{3/(1+2\pi\tilde{\lambda}\tilde{v}^4)}$ is the typical 
scale 
of the CEH, $\tilde{\lambda}_{Comp}
= \sqrt{2/\tilde{\lambda}\tilde{v}^2}$ is the typical
size of the structure and $n$ is the node number of the 
scalar
field\cite{TT1}.
If $\tilde{\lambda}$ becomes small, the size of the
structure becomes large compared to the size of the CEH.
Below the critical parameter $\tilde{\lambda}_{cr}$ the
structure can not be packed into the radius 
$\tilde{R}_{cos}$ and the
non-trivial solution disappears. 
Let us consider the effect of the existence of the BEH to 
the
critical parameter. Rescaling $\tilde{r}$ and $\tilde{m}$ by 
the
BEH radius $\tilde{r}_B$ as $\bar{r} \equiv 
\tilde{r}/\tilde{r}_B$ 
and $\bar{m} \equiv \tilde{m}/\tilde{m}_B$, and taking an
approximation $\tilde{\phi} \gg \tilde{\phi}^{\prime} 
\approx 0$, 
i.e., near the critical parameter
$\tilde{\lambda}_{min}$, the field equations become
\begin{eqnarray}
& & \frac{d\bar{m}}{d\bar{r}}  \approx  
\tilde{\lambda}\tilde{r}_B^2
    \bar{r}^2 (\tilde{\phi}^2-\tilde{v}^2)^2,
\\
& & \frac{d\bar{\delta}}{d\bar{r}}  \approx  0,
\\
& & \tilde{\lambda}\tilde{r}_B^2
   \tilde{\phi} (\tilde{\phi}^2-\tilde{v}^2)   \approx  0,
\end{eqnarray}
where $\tilde{\lambda}\tilde{r}_B^2$ can be considered as 
the 
effective self-coupling constant and the critical
solutions are controlled by it. Consequently 
$\tilde{\lambda}_{min}$
becomes small when we consider the large BEH radius.

On the other hand 
$\tilde{\lambda}_{max}$ is a new
restriction which can not be seen in the regular
solution case. From Fig.~3 the maximum BEH radius
for each $\tilde{\lambda}$ branch becomes small as 
$\tilde{\lambda}$ increases. Hence if we fix the BEH radius, the 
non-trivial solution disappears at a
certain critical value of $\tilde{\lambda}$. 
This is a $\tilde{\lambda}_{max}$
curve. The types of the critical solutions 
are different depending
on $\tilde{\lambda}_{max}$ or equivalently 
$\tilde{v}_{max} \equiv \tilde{v}(\tilde{\lambda}
=\tilde{\lambda}_{max})$. When
$\tilde{v}_{max}  
\mbox{\raisebox{-1.ex}{$\stackrel
     {\textstyle>}{\textstyle\sim}$}} 0.14$
for $\tilde{r}_{B}=0.6$ the critical solution is the
extremal solution. Otherwise
the critical solution is not the extremal
one, but is shown by the points plotted on the
solution for $\tilde{\lambda} = 2000$, $5000$
in Fig.~3.

%
%
%

\section{Stability Analysis}
\label{sec:Stability}

In the previous section we found  black hole solutions with
scalar
hair in de Sitter spacetime although there is no counterpart
in asymptotically flat spacetime. This means that the no-hair
conjecture may not hold in  asymptotic de Sitter 
spacetime.
In this section we investigate the stability
of new solutions by using a linear perturbation method in 
order
to check whether the scalar hair is really physical or 
not.
Here we focus only on the radial modes.
 
First we expand the field functions around the static 
solution
$\tilde{\phi}_{0}$, $\tilde{m}_{0}$ and $\delta_{0}$ as 
follows:
\begin{eqnarray}
\tilde{\phi} (\tilde{t}, \tilde{r}) & = & \tilde{\phi}_{0} 
(\tilde{t})
   + \frac{\tilde{\phi}_1 (\tilde{t}, \tilde{r})}{\tilde{r}} 
\epsilon, 
\\
\tilde{m} (\tilde{t}, \tilde{r}) & = & \tilde{m}_{0} 
(\tilde{t})
   + \tilde{m}_1(\tilde{t}, \tilde{r}) \epsilon, 
\\
\delta (\tilde{t}, \tilde{r}) & = & \delta_{0} (\tilde{t})
   + \delta_1(\tilde{t}, \tilde{r}) \epsilon.
\end{eqnarray}
Here $\epsilon$ is an infinitesimal parameter.
Substituting them into the field functions 
(\ref{beq1}) $\sim$ (\ref{beq5}) and dropping the second
and higher order terms of $\epsilon$,
we find
\begin{equation}
\dot{\tilde{m}}_1 = 4\pi \tilde{r}^2 
f_0 \tilde{\phi}_{0}^{\prime} \dot{\tilde{\phi}}_1,
\label{lieq1}
\end{equation}
\begin{eqnarray}
-e^{\delta_{0}} f_{0}^{-1} \ddot{\tilde{\phi}}_{1}
+\left[ e^{-\delta_{0}} f_{0} 
            \tilde{\phi}_{1}^{\prime}\right]^{\prime}
& - & \left[\frac1{\tilde{r}} \left( e^{-\delta_{0}} f_{0}
   \right)^{\prime}
   +8\pi \tilde{r} e^{-\delta_{0}} \lambda 
(\tilde{\phi}_{0}^{2} 
   -\tilde{v}^{2}) 
   \tilde{\phi}_{0}\tilde{\phi}_{0}^{\prime}
   +   \lambda e^{-\delta_{0}}(3\tilde{\phi}_{0}^{2} 
  -\tilde{v}^{2}) 
   \right] \tilde{\phi}_{1}  \nonumber \\
& & -\left[\frac2{\tilde{r}} \left( \tilde{r} 
e^{-\delta_{0}} 
   \tilde{\phi}_{0}^{\prime} 
   \right)^{\prime}
   -8\pi \tilde{r} e^{-\delta_{0}} 
\tilde{\phi}_{0}^{\prime 3}
   \right] \tilde{m}_{1} =0,
\label{leq2}
\end{eqnarray}
where $f_{0} = 1-2\tilde{m}_{0}/\tilde{r}-1/3$.
Next we set 
$\tilde{\phi}_{1}  =  \xi(\tilde{r}) 
e^{i\tilde{\sigma} \tilde{t}}$
and
$\tilde{m}_{1} =  \eta(\tilde{r}) 
e^{i \tilde{\sigma} \tilde{t}}$.
If $\tilde{\sigma}$ is real, $\phi$ oscillates around the 
static
solution and then the solution is stable. On the other hand, 
if
the imaginary part of $\tilde{\sigma}$ is negative, the
perturbation
$\tilde{\phi}_1$ and $\tilde{m}_1$ diverges exponentially 
with time and then the
solution is unstable.
By Eq. (\ref{lieq1}) the relation between $\xi$ and $\eta$ 
is
$\eta = 4\pi \tilde{r} f_{0} \tilde{\phi}_{0}^{\prime} \xi$.
Then the perturbation equation
of the scalar field becomes
\begin{equation}
-\frac{d^{2}\xi}{d\tilde{r}^{\ast 2}} + \tilde{U}(\tilde{r}) 
\xi 
            = \tilde{\sigma}^{2} \xi,
\label{leq4}
\end{equation}
where we employ the tortoise coordinate $\tilde{r}^{\ast}$ 
defined
by
\begin{equation}
\frac{d\tilde{r}^{\ast}}{d\tilde{r}}  = 
\frac{e^{\delta_{0}}}{f_{0}},
\end{equation}
and the potential function is
\begin{eqnarray}
\tilde{U}(r)  = & & e^{-\delta_{0}}f_{0}
  \left[\frac1{\tilde{r}} \left( e^{-\delta_{0}} 
f_{0}\right)^{\prime}
   +8\pi \tilde{r} e^{-\delta_{0}} \lambda 
   (\tilde{\phi}_{0}^{2} -\tilde{v}^{2}) 
   \tilde{\phi}_{0}\tilde{\phi}_{0}^{\prime}
   +   \lambda e^{-\delta_{0}}
   (3\tilde{\phi}_{0}^{2} -\tilde{v}^{2}) \right.
 \nonumber \\
 & & \left. +  4\pi \tilde{r}f_{0}\tilde{\phi}_{0}^{\prime}
   \left\{ \frac2{\tilde{r}} \left( \tilde{r} 
e^{-\delta_{0}} 
   \tilde{\phi}_{0}^{\prime} 
   \right)^{\prime}
   -8\pi \tilde{r} e^{-\delta_{0}} 
\tilde{\phi}_{0}^{\prime 3}
   \right\} \right].
\end{eqnarray}
Fig. 5(a) shows the potential functions $U(r)$ of the
solution with $\tilde{\lambda}=300$, $\tilde{v}=0.1$
and $n=1$.
 
Since $d^2\xi / d\tilde{r}^{\ast 2}=U(r)
=0$ on both horizons, $\xi $ must approach zero
as $\tilde{r}^{\ast} \to \pm \infty$
for the negative mode by the
regularity  of Eq. (\ref{leq4}). Under this boundary
condition we have searched for the negative eigenmodes
and found them as shown  in Fig.~5(b). 
These modes are bound states ($m=0$) and there is no excited 
mode ($m=1$)
for $n=1$. For the solution with $n=2$, however,
there are two negative modes.
We expect that the node number of the scalar field of the 
static solution $n$
exactly corresponds to the number of their
negative modes $m$. Fig.~5(c) gives the eigenvalues. We also
plot the eigenvalues of the excited Schwarzschild-de Sitter
case with dashed lines for comparison. At the point $P$,
where the new solution branch coincides with
the Schwarzschild-de Sitter branch, another
unstable mode, which is an excited mode, appears in the
Schwarzschild-de Sitter branch.
This result is consistent with the analysis using
catastrophe theory\cite{Thom,TT2}, and this stability
change is classified into
the swallow tail catastrophe.
Although there is the
$r_B \to 0$ limit  for large $\tilde{\lambda}$ as 
we mentioned, the eigenvalue 
does not vary 
continuously to that of the regular solution because of the
difference of the boundary condition around the origin.
Varying the parameters we
found negative eigenmodes for every non-trivial solution.
As a result, all of the new solutions are unstable, even
against the radial perturbations and the scalar
hair falls out easily.
Thus, although we found black hole solutions with
scalar hair in the presence of cosmological
constant, such hair is not physical and the no-hair
conjecture seems to hold even in the de Sitter spacetime.

%
%
%

\section{Conclusion}
\label{sec:Conclusion}


We examined the no-hair conjecture in the
presence of a cosmological constant. For the first
step,  the real scalar field was considered as the matter
field and the spacetime was assumed to be static spherically
symmetric. When the scalar field is massless or has a convex 
potential like mass term, it was proved that there is no
regular black hole solution. However we can not find any 
proof
excluding  scalar hair in the general positive potential 
case.
Therefore we searched for black hole solutions which have a 
scalar
field with a double well potential, and found them by
numerical calculations. Their field configurations spread
out to the cosmological scale and are classified by the
node number of the scalar field. For large BEH a solution
branch ends up with the extreme solution in the 
$\tilde{M}$-$\tilde{r}_B$ diagram, while for
small BEH the behavior is different, depending on the self-
coupling
constant $\tilde{\lambda}$. If $\tilde{\lambda}$ is large, 
we can 
take the limit $\tilde{r}_B \to 0$ and the corresponding solution
becomes regular a solution without the BEH. If $\tilde{\lambda}$ 
is
small, the new solution branch hits the excited 
Schwarzschild-de Sitter branch
at a non-zero BEH radius. The new solutions have critical 
parameters $\tilde{\lambda}_{min}$ and 
$\tilde{\lambda}_{max}$
between which there are  non-trivial solutions.
$\tilde{\lambda}_{min}$ is determined by the ratio of the 
size 
of the structure to the size of the
CEH, while $\tilde{\lambda}_{max}$ comes from the extremes
of the solutions.

In order to specify whether the scalar hair we found is 
physical
or not, we investigated the stability of new solutions 
by
using a linear perturbation method.  As a result all of the new 
solutions
have negative eigenmodes and were found to be unstable. Thus
the scalar hair is not real but a wig which falls off 
easily.
Although we have considered only one real scalar field with 
a double well potential, we expect that the general no-scalar hair 
conjecture
holds even if the cosmological constant exists.

Since these new solutions are unstable, the scalar field
will be swallowed by the black hole and/or will escape to 
infinity
over CEH. Then the solution becomes the stable
Schwarzschild-de Sitter solution (a) with the same or
smaller mass than the initial one. It is interesting to
consider the development of the solution with maximum mass
$\tilde{M}=1/3$. 
Although a part of the scalar field escapes to infinity in the
general case, it would be possible to 
set up the initial data where  all the
energy of the scalar field collapses into the black hole. 
If the third law of  
black hole thermodynamics is valid even in the present case,
it should take an infinite amount of time for the black hole 
to swallow the scalar
field and for the BEH to become degenerate with the 
CEH. It has been shown that 
similar phenomena occur in the
evolution of the Kastro-Traschen solution\cite{KT}, which
can be interpreted as black holes with $Q=M$ are
balanced to each other in CEH. For small black holes, they
can coalesce to form a larger black hole. However 
when the size of the black holes is larger than a
certain critical radius,
coalescence does not occur even if we set the initial data 
to give the
black holes a large initial velocity toward each 
other\cite{KT2}.
Our case would correspond to the critical case that the 
resultant
BEH radius coincides with CEH.

As for the rotating case, the no-scalar hair conjecture can not 
been
proved by using the techniques adopted in 
Ref.~\cite{Bek,Bek2,Sud,Heu} even in the asymptotically flat
case because they strongly depend on spacetime symmetry,
i.e., staticity and spherical symmetry. On the other hand
there is the beautiful result of the uniqueness theorem in the 
asymptotically flat case\cite{BHT,Cha,Bek,Har,Tei}. 
It seems worth  investigating
its counterpart in the asymptotic de Sitter case.
At a first glance, however, we will soon find that the
cosmological constant prevents us from constructing Ernst
type equations and that different approaches are needed.
We leave them open questions.


\vspace{.7cm}

-- Acknowledgments --

We would like to thank 
Dmitri V. Gal'tsov, Akio Hosoya, Hideki Ishihara and Kei-
ichi Maeda 
for useful discussions and
Julian McKenzie-Smith for his critical reading of our paper. 
This work was supported partially by the Grant-in-Aid
for Scientific Research Fund of the Ministry of Education,
Science and Culture (T.T. and K.M.),
by the Grant-in-Aid for JSPS 
(No. 199704162(T.T.) and 199605200(K.M.)).


\newpage
\begin{flushleft}
{Figure Captions}
\end{flushleft}


\noindent
\parbox[t]{2cm}{ FIG. 1:\\~}\ \
\parbox[t]{14cm}
{The configurations of the scalar field 
$\tilde{\phi}=\sqrt{G}\phi$ with
one node ($n=1$). We set $\tilde{v}=\sqrt{G}v=0.1$,
$\tilde{\lambda}=\lambda/G\Lambda=700$
and 
show the solutions
for $\tilde{r}_B=\sqrt{\Lambda} r_B=0.2$, $0.4$, $0.6$, 
$0.8$. 
We also plot the regular solution without BEH by the
dashed line for comparison. The left and right end
points of each lines are BEH and CEH, respectively, and
a dotted line traces them.
As the BEH radius becomes large, the BEH coincides with 
the CEH and the solution becomes extremal.
}\\[1em]
\noindent
\parbox[t]{2cm}{ FIG. 2:\\~}\ \
\parbox[t]{14cm}
{The configurations of the scalar field 
$\tilde{\phi}=\sqrt{G}\phi$ with
one node ($n=1$). We set $\tilde{v}=\sqrt{G}v=0.1$, 
$\tilde{r}_B=\sqrt{\Lambda} r_B=0.2$
and show the solutions
for $\tilde{\lambda}=\lambda/G\Lambda=400$, $700$, 
$2000$, $5000$. 
As $\tilde{\lambda}$ becomes small, $\tilde{\phi}$ 
approaches
zero and the solution coincides with the excited 
Schwarzschild-de 
Sitter 
solution (b) at $\tilde{\lambda} = \tilde{\lambda}_{min}$.
}\\[1em]
\noindent
\parbox[t]{2cm}{ FIG. 3:\\~}\ \
\parbox[t]{14cm}
{The mass-BEH radius diagram of the new solutions.
We set $\tilde{v}=\sqrt{G}v=0.1$
and show the solutions
for $\tilde{\lambda}=\lambda/G\Lambda=300$, $400$, $700$, 
$2000$, $5000$.
The dashed and dotted lines are the Schwarzschild-de Sitter 
solutions
(a) and the excited Schwarzschild-de Sitter solutions (b), 
respectively.
For small $\tilde{\lambda}$ there is a lower limit of the 
BEH, while
we can take the BEH zero limit for the solution with large 
$\tilde{\lambda}$. The new solution branch and the exited
Schwarzschild-de Sitter branch approach each other for large 
BEH and finally coincide. At this point the BEH and CEH 
degenerate
and the solution becomes extremal.
}\\[1em]
\noindent
\parbox[t]{2cm}{ FIG. 4:\\~}\ \
\parbox[t]{14cm}
{The critical parameters of the non-trivial solutions with 
$\tilde{r}_B=\sqrt{\Lambda} r_B=0.6$ and
$n=1$. Between $\tilde{\lambda}_{max}$ and 
$\tilde{\lambda}_{min}$ 
curves there exist  non-trivial solutions.
We also plot the critical parameters of the regular solution 
by the dashed line for comparison. 
On the critical lines $\tilde{\lambda}_{mim}$ the non-
trivial solutions
coincide with the excited Schwarzschild-de Sitter solution 
(b).
On the other hand the non-trivial solutions become extremal 
on the
critical lines $\tilde{\lambda}_{max}$ with 
$\tilde{v}\mbox{\raisebox{-1.ex}{$\stackrel
{\textstyle>}{\textstyle\sim}$}} 0.14$,
otherwise they correspond to the 
solution dotted in Fig.~3.
}\\[1em]
\noindent
\parbox[t]{2cm}{ FIG. 5:\\~}\ \
\parbox[t]{14cm}
{(a) The configurations of the potential function $U$ of 
the
linear perturbation equation for 
$\tilde{v}=\sqrt{G}v=0.1$, 
$\tilde{\lambda}=\lambda/G\Lambda=300$, $n=1$
and $\tilde{r_B}=\sqrt{\Lambda} r_B=0.4$, $0.6$, $0.8$.
(b) The configurations of eigenmodes of the perturbation
equation with the same parameters as (a).  We find the
only bound state ($m=1$).
The eigenvalue for each mode is 
$\tilde{\sigma}^2=-0.4165$, $-0.1522$, $-0.0298$, 
respectively.
(c) The eigenvalue of the linear perturbation equation
for $\tilde{v}=\sqrt{G}v=0.1$, 
$\tilde{\lambda}=\lambda/G\Lambda=300$ and $n=1$.
Dotted lines are those of the excited Schwarzschild-de
Sitter solution. At the point P where the non-trivial 
solution
disappears, the excited mode of the Schwarzschild-de
Sitter solution ($m=1$) appears. This is consistent with the
analysis made using catastrophe theory.

}\\[1em]

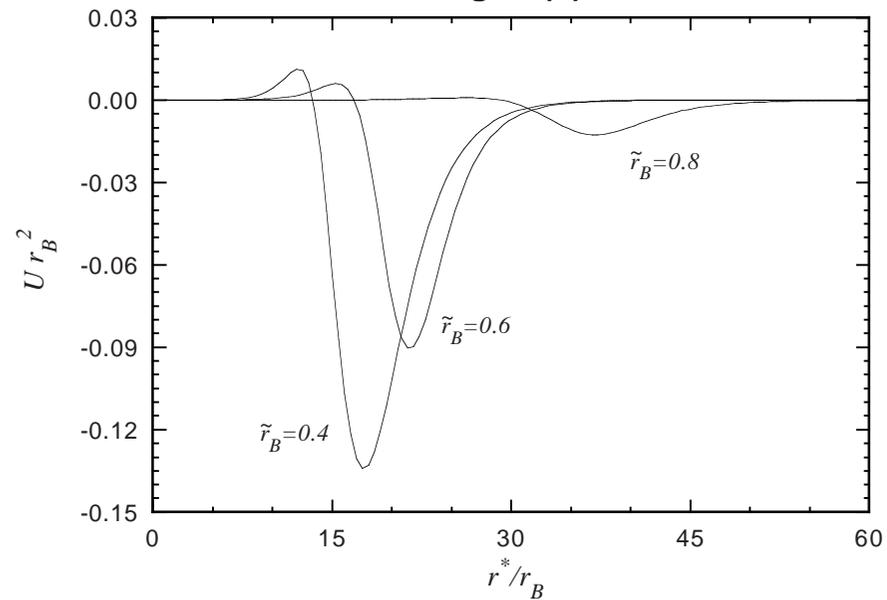
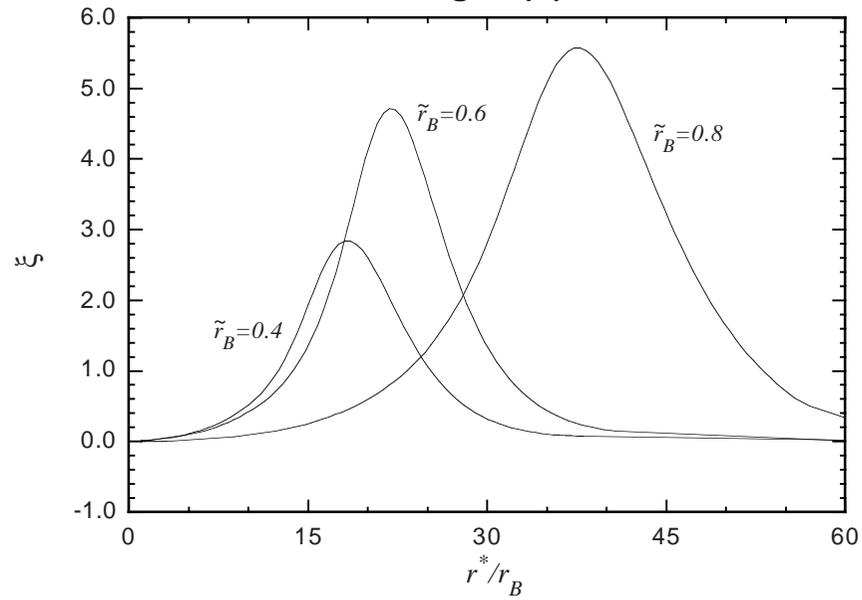
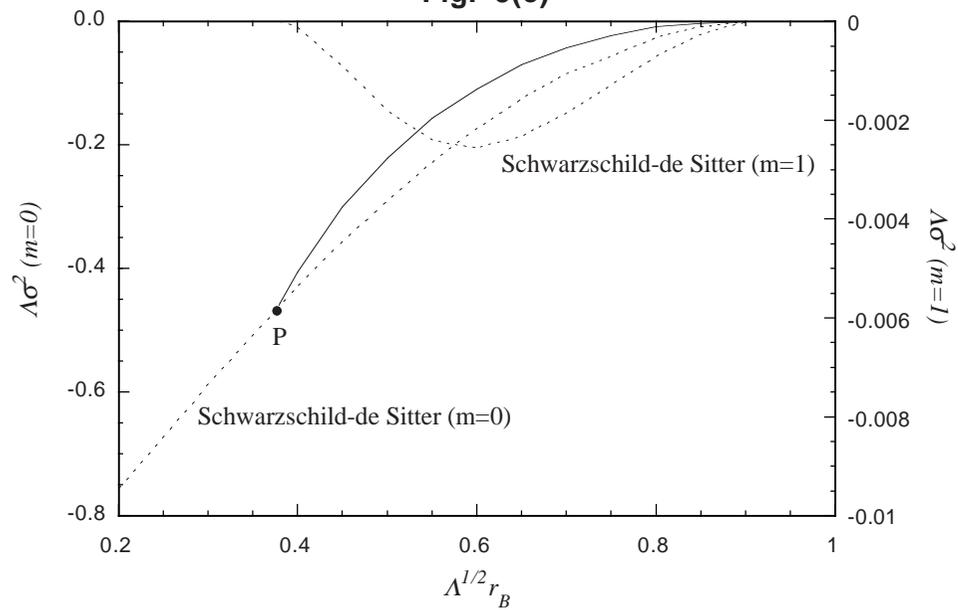
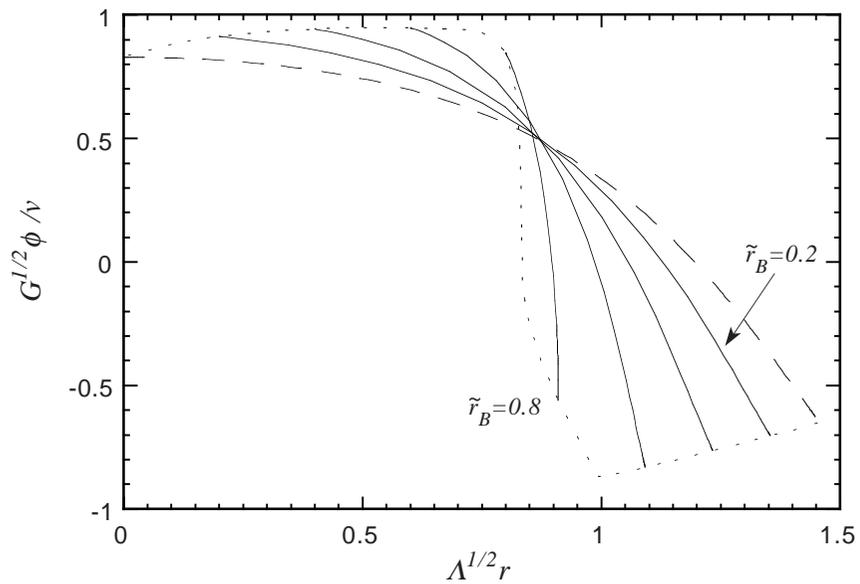
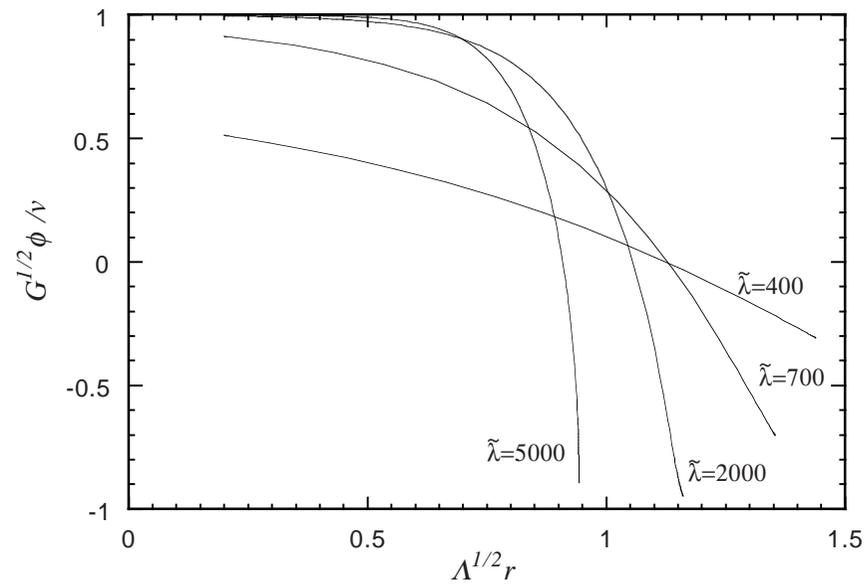
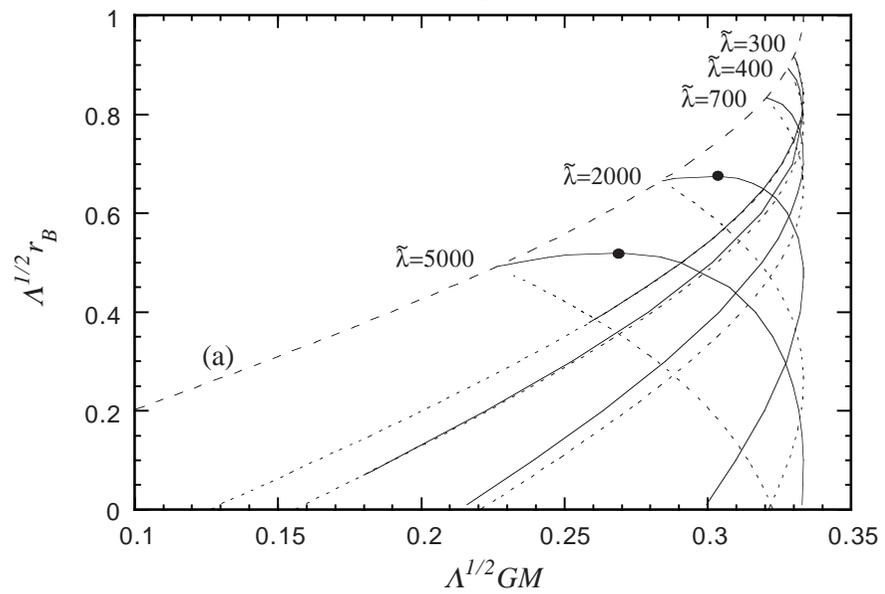
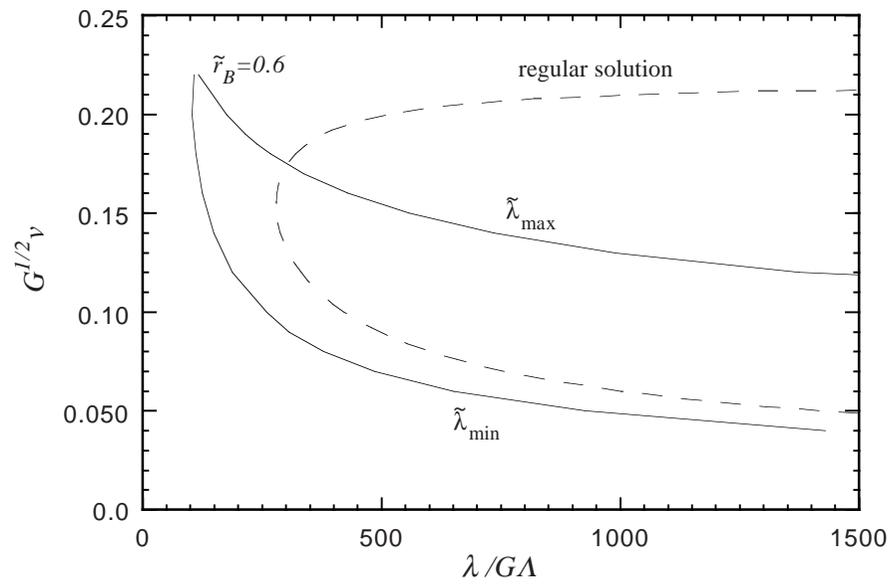
\end{document}